# Direct observation of geometric and sliding ferroelectricity in amphidynamic crystal

Le-Ping Miao[1,2,3,6], Ning Ding[4,6], Na Wang[2,6], Chao Shi[2], Heng-Yun Ye[2], Linglong Li[4], Ye-Feng Yao[5], Shuai Dong[4] & Yi Zhang[1,2,3]

[1] Institute for Science and Applications of Molecular Ferroelectrics, Key Laboratory of the Ministry of Education for Advanced Catalysis Materials, Zhejiang Normal University, Jinhua 321004, China

[2] Chaotic Matter Science Research Center, Department of Materials, Metallurgy and Chemistry & Jiangxi Provincial Key Laboratory of Functional Molecular Materials Chemistry, Jiangxi University of Science and Technology, Ganzhou 341000, China

[3] Jiangsu Key Laboratory for Science and Applications of Molecular Ferroelectrics, Southeast University, Nanjing 211189, China

[4] School of Physics, Southeast University, Nanjing 211189, China

[5] Shanghai Key Laboratory of Magnetic Resonance, School of Physics and Materials Science, East China Normal University, Shanghai 200062, China

[6] These authors contributed equally: Le-Ping Miao, Ning Ding, Na Wang.

Correspondence and requests for materials should be addressed to Y.Z. (email: yizhang1980@seu.edu.cn) or to S.D. (email: sdong@seu.edu.cn).

**ABSTRACT**

Sliding ferroelectricity was recently theoretically proposed and then experimentally confirmed as a new kind of polarity existing in two-dimensional materials. However, due to their weak polarization and poor electrical insulation, all available experimental evidences till now are indirect, most of which are based on the transport properties in nanoscale or piezoresponse force microscopy. Here, we report the direct observation of sliding ferroelectricity in an amphidynamic crystal (15-Crown-5)$Cd_3Cl_6$. This coordination polymer is a van der Waals material, which is constructed by inorganic stators and organic rotators. After the freezing of rotators, an electric dipole is generated in each layer driven by the geometric mechanism, meanwhile a comparable ferroelectric polarization originates from the interlayer sliding. The net polarization of these two components can be directly measured and manipulated. Our finding provides new insight into low-dimensional ferroelectrics, especially the controlling of synchronous dynamics of rotating molecules and sliding layers in solids.

Ferroelectric materials are technologically important smart materials, which can facilitate the conversions among electric, magnetic, mechanical, thermal, and optical signals. The most commercially popular ferroelectrics are three-dimensional (3D) inorganic crystals, such as lead zirconate titanate and barium titanate. With the purpose of more energy-saving, higher integration, and flexible applications, alternatives of unconventional ferroelectrics become urgently attractive. For example, two-dimensional (2D) ferroelectrics are rapidly growing to form an emerging branch of polar systems[1-3], with several experimentally confirmed materials, such as SnTe monolayer[4], $CuInP_2S_6$ few layers[5], and α-$In_2Se_3$ nanoflakes[6], and more theoretical predicted ones. In these materials, ferroelectricity can persist till the ultra-thin limit, e.g., atomic level, superior than conventional 3D perovskites. In addition, exotic scientific effects and new physical phenomena have been revealed in these systems, such as giant negative piezoelectricity and sliding ferroelectricity[7,8].

Sliding ferroelectricity is a new type of polarity, whose origin is unique in van der Waals (vdW) materials. Concretely, the stacking mode of vdW layers breaks the inversion symmetry, generating out-of-plane polarizations. Soon after the theoretical prediction by Wu *et al*[9], following experiments found evidences of sliding ferroelectricity in $WTe_2$ bilayer/few-layer/ flakes[10-12], bilayer *h*-BN[13,14], bilayer 1*T'*-$ReS_2$[15], and bilayer *R*-$MX_2$ (*M*=Mo/W, *X*=S/Se)[16]. And this sliding concept has also been generalized to broader scopes, such as Moiré ferroelectricity[8] and intralayer sliding[17]. Despite these achievements, currently all available experimental evidences are indirect (as summarized in Supplementary Table 1). In fact, many of these materials are semiconductors or even semi-metals (e.g. $WTe_2$), which make the standard electrical measurements of their ferroelectricity challenging. Alternatively, their ferroelectricity were mostly characterized by the transport behaviors in nano-devices or piezoresponse force microscopy (PFM), with inevitable interference from leakage currents and environments. For example, the coercive electrical field observed in transport measurements is high (in the order of 0.1 V/nm)[10,12], instead of the expected small value according to its sliding mechanism. Large coercivity and serious leakage will corrode the technical value of ferroelectrics.

Another interesting branch of polar materials is the family of molecular ferroelectrics[18,19]. The most effective way to produce polarity in these materials is to introduce dipolar molecules into lattices, e.g. pyridinium[20], benzylammonium[21], diisopropylammonium[22], pyrrolidinium[23], cyclohexylammonium[24]. Those dipolar molecules are dynamically disordered at high temperatures and ordered below critical temperatures ($T_C$'s), leading to the alignment of dipoles and spontaneous polarizations. The ferroelectricity in a series of plastic molecular ferroelectrics including molecular perovskite ferroelectrics is owned to this method[25-27].

Besides dipolar dynamical molecules, there are a few non-dipolar dynamical molecules, such as crown-ethers and molecular gyroscopes. For example, in a few crown-ether-based ferroelectrics, such as [$(C_7H_{10}NO)$(18-Crown-6)][$BF_4$], [$(C_7H_{10}NO)$(18-Crown-6)][$ReO_4$], and (m-FAni)-(DB 18-Crown-6)[$Ni(dmit)_2$]$_{10}$ (m-FAni= m-fluoroanilinium; $dmit^{2-}$= 2-thioxo-1,3-dithiole-4,5-dithiolate)[28-30], ferroelectricity is assigned to the alignment of the rotating p-methoxyanilinium and m-fluoroanilinium, whereas the crown-ethers act as stators. Compared with the dipolar molecules, these non-dipolar dynamical molecules are relatively ignored and thus not fully understood regarding their ferroelectricity.

In this work, the ferroelectricity in an amphidynamic vdW crystal, coordination polymer (15-Crown-5)$Cd_3Cl_6$ (CCC)[31], is unambiguously confirmed. Interestingly, its ferroelectricity owns dual sources. The first contribution comes from individual layers, whose dipoles are driven by the geometric frustration of oxygen ions, once the rotational dynamics of crown-ethers are frozen. The second contribution is generated by the interlayer sliding, which is the dominant component of net polarization. These two components are coupled, going beyond the established scenarios. Most importantly, it is decisive to directly prove the sliding ferroelectricity in this highly-insulating crystal, which extends the sliding ferroelectricity to broader scopes and reduces the technical difficulties to manipulate sliding ferroelectricity.

**RESULTS**

**Proof of amphidynamic crystal**

Colorless and transparent rod-like single crystals of CCC with size 3×3×8 mm$^3$ (Fig. 1a and Supplementary Fig. 1) were grown by slow evaporation of a clear methanol solution containing stoichiometric amounts of $CdCl_2$ and 15-Crown-5 at room temperature. Phase purity was confirmed by powder X-ray diffraction and infrared radiation (IR) analysis, as shown in Supplementary Figs. 2-3.

As revealed by the differential scanning calorimetry (DSC) measurement, CCC undergoes a phase transition at $T_C$=320 K (Supplementary Fig. 4). Its crystal structures are determined by the single-crystal X-ray diffraction at 253−343 K (Supplementary Note 1 & Table 2). The high-temperature phase (HTP) is centrosymmetric with space group $P2_1/n$, while the symmetry of the low-temperature phase (LTP) is lowered to space group $P2_1$. Despite this difference, the crystal frameworks of both phases are similar. As shown in Figs. 1b, c, its unit cell (u.c.) consists of two layers (A and B) of adducts stacking along the $b$-axis, and its structural unit can be described as a molecular pinwheel, where the crown-ether acts as the rotator.

There are two kinds of Cd ions in different chemical environments. The first kind of Cd ion is surrounded by five Cl ions, giving the coordination geometry of tetragonal pyramid. The neighboring tetragonal pyramids are edge-shared, forming one-dimensional chains. The second kind of Cd ion sites at the center of 15-Crown-5, coordinated by five O ions in the equatorial plane and two Cl ions in the axial positions, i.e., with the coordination geometry of pentagonal bipyramid. The pentagonal bipyramids link the chains into layers by sharing the vertexes with tetragonal pyramids. These layers are electrically neutral and accordingly packed by vdW forces with the inter-layer distance of 7.305(3) Å at 343 K (Supplementary Fig. 5).

In the HTP, the second kind Cd ion is located at the inversion center (Fig. 1d). The model of the 15-Crown-5 determined from the Fourier difference map has five O atoms distributing over ten sites and the ten C atoms showing obvious larger thermal ellipsoids elongate (Supplementary Fig. 6). These pieces of evidence initially reveal the rotation of 15-Crown-5 in the HTP, which are frozen to the fixed positions in the LTP (Fig. 1e).

To reveal the mechanical movement of the crown-ether, solid-state nuclear magnetic resonance (NMR) spectroscopy has been employed, which in principle can probe the molecular motions in the frequency range from Hz to MHz. Here, the motionally modulated $^{13}$C isotropic signal is shown in Fig. 1f. In the spectra, only one signal centered at 70.7 ppm was observed, which can be assigned to the $CH_2$ groups of crown-ether. At 250 K, the signal shows a large width with a half-width of 1250 Hz. This signal becomes narrower and narrower upon heating. At

340 K, the signal half-width is reduced to only 70 Hz. This signal narrowing phenomenon indicates that the motions of crown-ether gradually increase with increasing temperature[32,33].

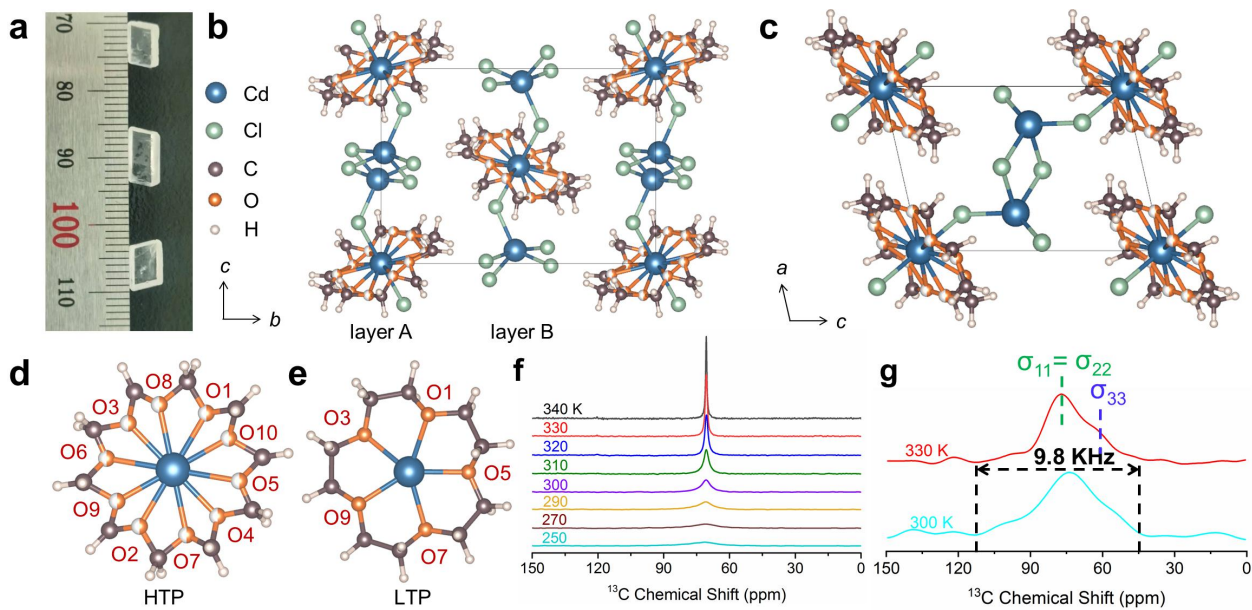

**Fig. 1 | Structure of (15-Crown-5)Cd₃Cl₆ (CCC). a**, The image of single crystals. **b,c**, Schematic of its atomic structure in the HTP. **b**, Side view. The vdW stacking is along the *b*-axis. **c**, Top view of each vdW layer. **d,e**, Top views of a crown-ether in the HTP and LTP, respectively. **f**, Temperature-dependent solid state $^{13}$C NMR spectra. These spectra were acquired by using the single pulse excitation pulse sequence with a recycle delay of 2 s; **g**, The $^{13}$C CSA patterns at 300 K (below $T_C$) and 330 K (above $T_C$).

The $^{13}$C chemical shift anisotropy (CSA) signals above and below $T_C$ are compared in Fig. 1g. At 300 K, the $^{13}$C CSA pattern shows a complex asymmetric tensorial lineshape, implying very restricted mobility of the crown-ether molecules. In contrast, the $^{13}$C CSA pattern at 330 K shows the typical features of an axially symmetric tensor with three principle values $\sigma_{11}=\sigma_{22}\neq\sigma_{33}$, indicating that the crown-ether molecule undergoes a restricted anisotropic motion close to an axial rotation[34]. In addition, molecular motions having the correlation time in the spectral time-scale can cause the significant lineshape perturbation[35,36]. As shown in Fig. 1g, the frequency ranges of the patterns change from 10 kHz at 300 K to 3 kHz at 330 K, accompanied by a distinct change in the pattern lineshape. To have such a significant lineshape perturbation, the frequency of the molecular motion must be much higher than 10 kHz, suggesting that the related molecular motions have the correlation time in the order of μs or sub μs, which is well in line with the following observations in the dielectric measurements.

In short, CCC is an amphidynamic crystal, with inorganic $CdCl_2$ chains as the stators and organic crown-ethers as the rotators above $T_C$.

**Characterization of ferroelectricity**

According to the above variable-temperature single crystal structure analysis, the order-disorder transition of crown-ethers induces a nonpolar-polar transition of CCC crystal, which can be further characterized by the second harmonic generation (SHG) measurement on powder samples[37]. As expected, the SHG signal appears just below $T_C$=320 K and its intensity increases gradually upon cooling (Fig. 2a). At room temperature, the SHG intensity of CCC is 6 times that of quartz. More details of SHG measurements and comparisons to other materials can be found in Supplementary Information (SI) (Supplementary Note 2, Figs. 7,8, & Table 3).

Besides the SHG signal, frequency-dependent dielectric curves are measured as a function of temperature, which can provide further information of polarity[38]. As shown in Fig. 2b and Supplementary Figs. 9a,b, the dielectric constant measured along the b-axis ($\varepsilon'_b$) is larger than those along the a-/c-axes ($\varepsilon'_a/\varepsilon'_c$), especially near $T_C$. Such dielectric anisotropy agrees with its space group $P2_1$, which allows spontaneous polarity only along the b-axis. This dielectric anomaly is remarkable over the frequency range 500 Hz - 1 MHz. For low frequencies, its real part $\varepsilon'_b$ reaches a peak at $T_C$ (e.g., ~30 at 1 kHz, 6 times of the room-temperature one) and its temperature dependence is typical of the continuous phase transition. For higher frequencies, $\varepsilon'_b$ shows two broad maxima below and above $T_C$. The temperature dependence of its imaginary part $\varepsilon''_b$ reveals a sharp maximum whose value decreases with measuring frequency (Supplementary Fig. 9c). The spectra are similar to those observed for diglycine nitrate and $Ca_2Sr(C_2H_5CO_2)_6$[39], known as dielectric critical slowing down. This critical slowing down is the characteristic of ferroelectrics with order-disorder type continuous phase transition.

To derive more information, the Cole-Cole diagrams are analyzed at five different temperatures slightly above $T_C$ (Insert of Fig. 2b). All these fitted plots are close to semicircles which are expected for the ideal Debye relaxation model, where the dipoles are almost non-interacting. Analytically, the dielectric dispersion function can be described as:

$$\varepsilon_b(\omega) = \varepsilon_{b,\infty} + \frac{\varepsilon_{b,0} - \varepsilon_{b,\infty}}{1+(i\omega\tau)^{1-\alpha}}, \tag{1}$$

where $\varepsilon_{b,0}$ and $\varepsilon_{b,\infty}$ are the static and high-frequency dielectric constants, respectively; $\omega$ is the angular frequency of applied field, $\tau$ is the mean relaxation time, and $\alpha$ is the width parameter measuring Debye relaxation time ($\alpha=0$ corresponds to the ideal Debye relaxation). The fitted $\tau$ and $\alpha$ for CCC are summarized in Supplementary Table 4. The fitted $\alpha$ values are close to zero, indicating that the distribution of the relaxation times are very narrow and the observed dielectric relaxation processes are of the mono-dispersive Debye relaxation. The relaxation time $\tau$ ranges from $4.82\times10^{-6}$ s to $7.54\times10^{-7}$ s in the temperature range from 321 K to 327 K.

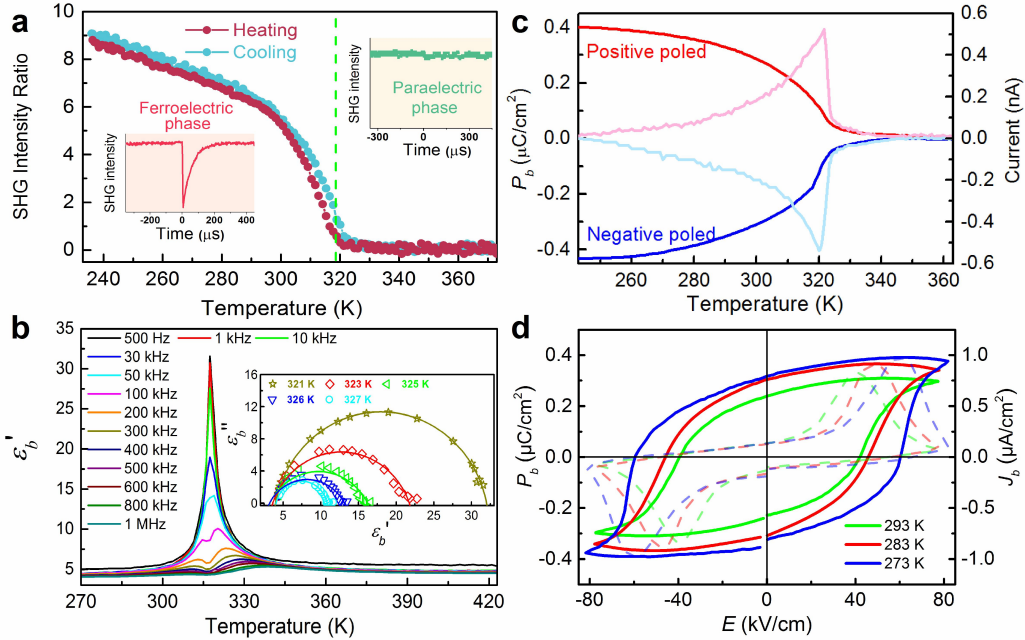

**Fig. 2 | Characterization of ferroelectricity. a**, The SHG intensity of CCC as a function of temperature on powder samples. The SHG intensity of quartz at room temperature is taken as the unit. Insert: the comparison of SHG signals above and below $T_C$. **b-d**, Electrical characterizations measured along the *b*-axis. **b**, The real part of dielectric constant ($\varepsilon_b'$) at different frequencies; Insert: the relationship between $\varepsilon_b'$ and $\varepsilon_b''$ at temperatures slightly higher than $T_C$, showing typical Cole-Cole diagrams. **c**, Pyroelectric current (light curves) and integrated polarization (dark curves), showing the emergence of polarization below $T_C$. **d**, The *J-E* (broken) and *P-E* (solid) curves, showing typical ferroelectric hysteresis loops.

To affirm its ferroelectricity, an electrical-switchable polarization below $T_C$ must be proved. Our measurement indeed shows a continuous curve of pyroelectric current below $T_C$ (peaked at $T_C$), whose sign can be switched by the poling field, as shown in Fig. 2c. These pyroelectric behaviors strongly support its ferroelectricity below $T_C$. Furthermore, a polarization-electric field (*P-E*) hysteresis loop is the most decisive evidence for ferroelectricity. To obtain the *P-E* loops, the current-field (*J-E*) curves are measured under different temperatures, as shown in Fig. 2d. There are two peaks due to charge displacement, indicating two stable states with opposite polarity. Variable-temperature *P-E* hysteresis loops are obtained according to the current integration (Fig. 2d). As temperature decreases from 293 K to 273 K, its coercive field enhances from 40 kV/cm to 60 kV/cm, namely the reverse of polarization becomes harder at lower temperatures. The saturated polarization ($P_S$) are in the range of 0.3-0.4 μC/cm$^2$ at 293-273 K, is in consistent with the pyroelectric values (Fig. 2c). Similarly, as the temperature is lower, the value of $P_S$ becomes larger, in qualitatively agreement with the increasing SHG signal intensity at low temperatures.

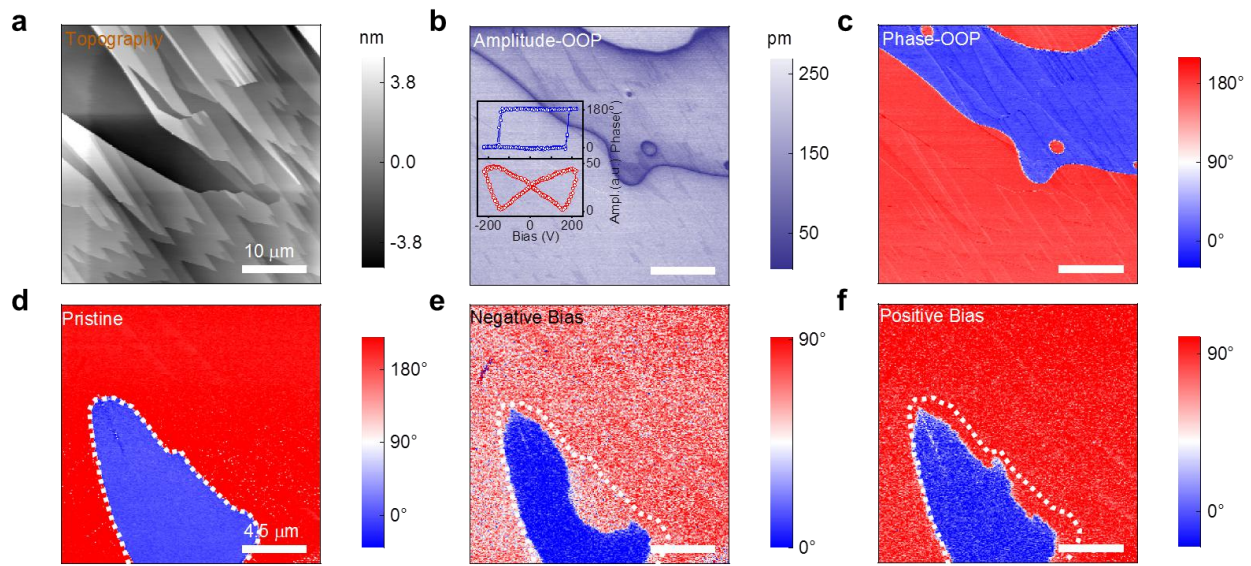

**Fig. 3 | PFM characterization and manipulation of ferroelectric domains. a**, Topology image of the crystal plane (010) of CCC. **b,c**, The corresponding vertical PFM amplitude and phase images. Insert of **b**, Phase-voltage hysteresis loop and amplitude-voltage butterfly loop. **d-f**, The electric writing of ferroelectric domain. From left to right: the initial phase image; after applying a negative bias voltage; after applying a positive bias voltage. The voltage ±220 V is applied on the central area for 20 s.

For ferroelectricity, the ferroelectric domain structure is another essential issue, which can be checked using PFM. Figure 3a is the topography of the (010) surface of CCC single crystal, and obvious domain textures from the amplitude and phase images in vertical PFM can be found in Figs. 3b,c. In contrast, in lateral PFM, very weak (and the same shape as the vertical one) domain textures of the amplitude and phase images are observed (Supplementary Figs. 10a-c). These evidences indicate that there are strong vertical piezoresponse and negligible lateral piezoresponse, implying that its polarization orientation is only along the *b*-axis.

To further reveal the switchable ferroelectric feature of CCC, the vertical PFM switch spectroscopy loop is measured, which is the relationship of phase and amplitude signal on DC tip bias on the same single crystal sample through the measurement of point-wise polarization (Insert of Fig. 3b). Besides, the switching of ferroelectric domain in the homogeneous PFM amplitude and phase signal area is also demonstrated, as shown in Figs. 3d-f (and Supplementary Figs. 10d-f). The ferroelectric domain can be reversed and switched back by applying bias voltages. Achieving the completing reversal of the domain reveals the ferroelectric nature of CCC.

**Geometric ferroelectricity**

Above experimental evidences have associated its ferroelectric polarization with the disorder-order transition of crown-ethers. However, the symmetry-breaking mechanism in CCC is different from most previously-studied amphidynamic crystals, namely here the crown-ethers play as the rotators, instead of the stators in many others cases where the attached asymmetric groups rotate[28-30].

The first mechanism involved here can be coined as the geometric ferroelectricity. In the HTP, the 10 oxygen sites (and 10 carbon, 20 hydrogen, and 2 chlorine ions) in each crown-ether own the inversion symmetry regarding to the central Cd ion. For example, the O1-Cd-O2 bond angle is 180° and the bond-lengths of Cd-O1 and Cd-O2 are equivalent (Fig. 1a). In the LTP, the five oxygen ions prefer to keep away from each other due to the Coulomb repulsion, i.e., to occupy the odd- or even-index sites (Fig. 1e). Then the inversion symmetry is naturally broken for each crown-ether, which further leads to stereo distortions of all ions in the crown-ether. As illustrated in Fig. 4a, all oxygen ions deviate from the crown-ether plane more or less (to keep away from each other), mainly in the -↑-↓- fluctuating manner (similar to the spin order in antiferromagnetic systems). Meanwhile, the peripheral H ions will align in the fluctuating manner of -↓-↓-↑-↑- correspondingly. The odd number of oxygen ions in each crown-ether frustrates this mode within a whole period, creating a soliton in each crown-ether, together with H ions in the peripheral side. The two apical Cl ions of crown-ether are also slightly influenced by the polar crown-ether, with staggered lengths of Cd-Cl bonds (slightly longer and shorter) and slightly bending Cl-Cd-Cl bond angle ~177.4° (Supplementary Fig. 11).

This mechanism is similar to the ferroelectric origin in hexagonal manganites/ferrites $R$MnO$_3$/$R$FeO$_3$ ($R$: rare earth or Y), where the displacements of $R$ ions due to structural trimerization are frustrated due to the triangular geometry[40]. Such geometric ferroelectricity belongs to the family of improper ferroelectricity, which is more robust than the proper ones.

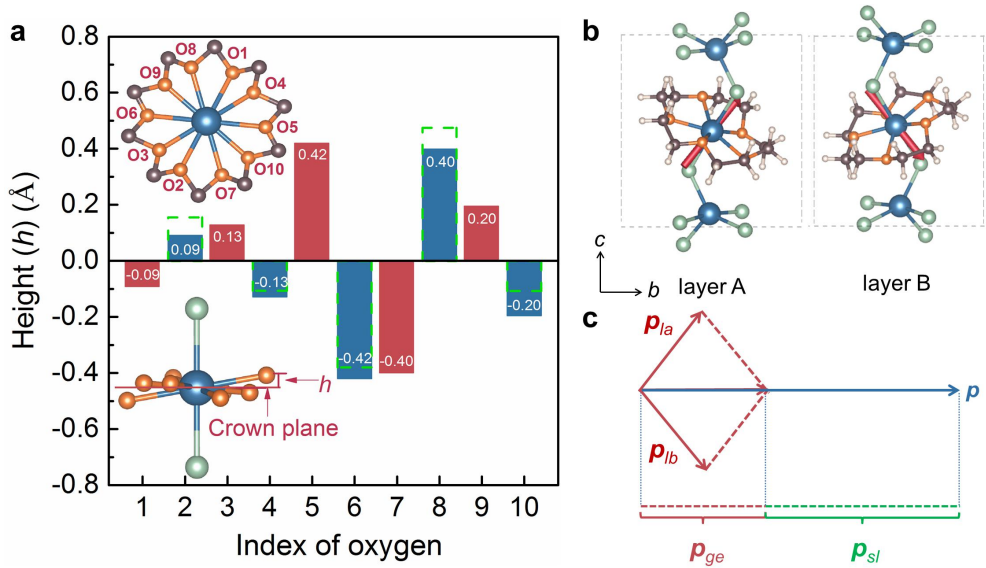

**Fig. 4 | Schematic of geometric ferroelectricity. a**, The height distribution of O ions within a crown-ether relative to the crown-ether plane, according to the experimental structures. Inserts: the definitions of index (upper) and height (lower). Solid bars with values: ten sites in the HTP; Broken bars: five sites in the LTP. Odd and even bars are distinguished by colours. In the HTP, the anti-symmetry exists between odd-index and even-index sites, which is broken in the LTP since only odd-index (or even-index) sites are occupied. A soliton of the -↑-↓- fluctuation is inevitable, which generates a dipole together with the peripheral C and H ions. **b**, The isolated layers A and B, rigidly extracted from the optimised bulk in the ferroelectric state. Their corresponding dipole vectors ($p_{la}$ & $p_{lb}$) are indicated as arrows. **c**, The superposition of $p_{la}$ & $p_{lb}$ leads to a net geometric dipole $p_{ge}$ along the $b$-axis, but it is smaller than the total dipole $p$ of a bulk u.c.. The surplus part is attributed to the sliding ferroelectric dipole, i.e., $p_{sl}$.

In short, the crown-ether in the LTP can generate a dipole for each individual layer, which can be generally expressed as ($p_a$, $p_b$, $p_c$) for layer A and (-$p_a$, $p_b$, -$p_c$) for layer B. This asymmetric (symmetric) relationship of the *ac*-component (*b*-component) between two layers are guaranteed by the $P2_1$ symmetry, which owns a two-fold screw axis and allows a net polarization is only along the *b*-axis. The existence of dipoles in individual layers and the relationship between layers can be confirmed using density functional theory (DFT) calculations (Supplementary Note 3), as summarized in Table 1 and illustrated in Fig. 4b.

**Table 1** | DFT dipoles (in unit of eÅ) for individual layers and bulk u.c.. The coordinate (*x*, *y*, *z*) is orthorhombic. The crystalline axes *a*∥*x*, *b*∥*y*, but *c* is along (-0.236, 0, 0.972).

|  | Dipole along (*x*, *y*, *z*) | Dipole along (*a*, *b*, *c*) |
|---|---|---|
| Isolated layer A | $p_{la}$ = (-0.085, 0.056, 0.270) | $p_{la}$ = (-0.0193, 0.056, 0.278) |
| Isolated layer B | $p_{lb}$ = (0.085, 0.056, -0.270) | $p_{lb}$ = (0.0193, 0.056, -0.278) |
| Isolated layers A+B | $p_{ge}$ = $p_{la}$ + $p_{lb}$ = (0, 0.112, 0) | |
| Bulk (A+B stacking) | $p$ = $p_{ge}$ + $p_{sl}$ = (0, 0.263, 0) | |

**Sliding ferroelectricity**

Although the geometric mechanism can qualitatively explain the ferroelectric polarization along the *b*-axis, the ferroelectric dipole of a bulk u.c. (0.263 eÅ) is obviously larger than the direct sum of two isolated layers (0.112 eÅ), as shown in Fig. 4c and Table 1, which is much beyond the allowable error of DFT calculations. Thus, other contribution must exist, which is non-negligible (in fact dominant) since its contribution reaches 57.4% of total.

This additional ferroelectric origin is from the vdW interlayer sliding. Although the experimental measurements of ferroelectricity can not directly distinguish the individual contributions of geometric one and sliding one, the existence of sliding ferroelectricity in CCC is unambiguously evidenced by monitoring ion positions, which can be determined from the single-crystal XRD measurements as well as the DFT structural relaxation. Careful analysis of its HTP and LTP structures indeed reveals the existence of interlayer sliding (Fig. 5a). Using the Cd ions in crown-ethers as the indication, the relative sliding distance in the *ac*-plane is 0.30 Å according to the experimental HTP and LTP structures, as confirmed in our DFT calculation: 0.62 Å between the +*P* and -*P* structures, where ±*P* denotes the ferroelectric states with opposite polarization. This magnitude of displacement is large enough to be precisely captured in DFT calculations. For reference, the DFT displacement of Mn in BaMnO$_3$ is only 0.038 Å[41].

As shown in Fig. 5b, the sliding indeed reduces the system energy to the minimum, and further sliding beyond the optimal level is not energy favorable. The polarization contribution due to interlayer sliding is perpendicular to the vdW layers, as expected for all sliding ferroelectrics, which superposes on the aforementioned geometric one along the *b*-axis (Fig. 5c). Naturally, the value of ferroelectric polarization is proportional to the sliding distance.

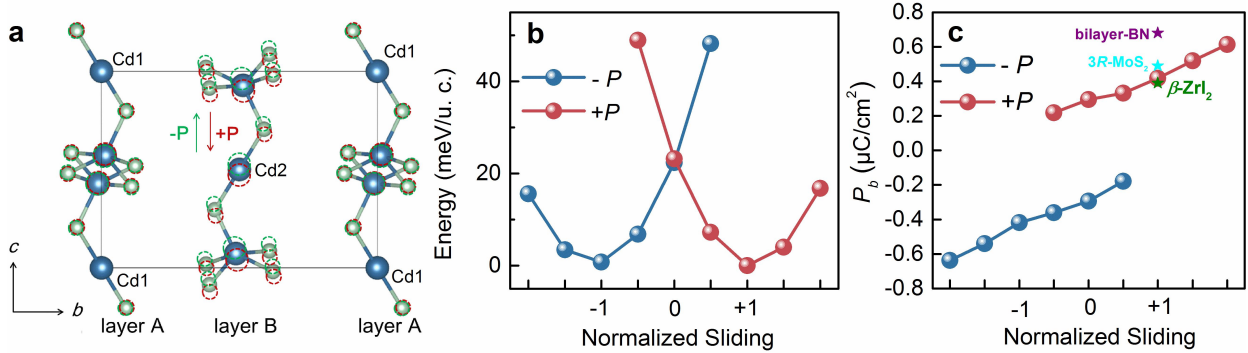

**Fig. 5 | Schematic of sliding ferroelectricity. a**, The ion displacements of $CdCl_2$ chains. Solid: HTP; Broken: LTP ($\pm P$). The organic parts are not shown for simplify. All displacements are defined relative to Cd1. **b,c**, The sliding-dependent energy and polarization. The sign of geometric contribution is distinguished by colours. The interlayer sliding is characterized by the relative shift of Cd2, normalized to its optimised one (see Supplementary Note 3 for more details). For comparison, the polarizations of other sliding ferroelectrics are shown as stars. The value of bilayer BN is from experiment[42], while others are DFT ones[8].

Different from previously studied sliding ferroelectrics, here nonzero ferroelectric polarizations exist in the so-called "0" states, partially from the geometric dipoles. In addition, although Cd2 is not sliding in the "0" states, other ions can still slide in a non-rigid manner (Supplementary Fig. 12) and generate a sliding dipole. Thus, even in the "0" states, the net polarizations are contributed by both the geometric and sliding sources. Futhermore, if the geometric contribution along the *b*-axis is turned off in our DFT calculation, the interlayer sliding and corresponding polarization along the *b*-axis also diminish to zero (Supplementary Note 3), implying the sliding one is probably secondary to the geometric one.

**DISCUSSION**

Due to the coupling between geometric and sliding components, the sliding is unidirectional upon the fixed sign of geometric polarization. It is reasonable since the existing of geometric ferroelectricity breaks the inversion symmetry. As a result, with a given geometric component, the energy profile as a function of sliding is no longer a double-well structure, but a single-well one. The expected double-well energy profile can be restored by flipping the geometric and sliding components together, satisfying the requirement of switchability of ferroelectricity. This coupling effect is absent in all previous sliding ferroelectrics with sole source of polarity, but similar to the case of hybrid improper ferroelectricity in $Ca_3Ti_2O_7$ and $Ca_3Mn_2O_7$[43-45], whose polarization also depends on two coupled degrees of freedom.

Limited by the relative weak vdW interaction, the polarizations of sliding ferroelectric may be naturally small (usually in the order of 1 $\mu C/cm^2$ or less)[8]. According to the perspective of current work, a promising direction for sliding ferroelectricity is to couple multiple ferroelectric sources in one material, which can strengthen the net polarization and provide more degrees of freedom for manipulation. For example, here the interlayer sliding can trigger the flipping of geometric polarization, and vice versa.

In addition, a noteworthy advantage of our system is its highly insulating property, which can reduce the leakage current in the ferroelectric measurements to a minimum level. Thus the direct ferroelectric measurements become possible and reliable, which are not easy jobs for those semiconducting or semi-metallic systems with sliding ferroelectricity. Indeed, our DFT calculation finds a band gap larger than 4 eV (Supplementary Fig. 13), in agreement with its transparent and colorless fact. As a result, our work can provide the direct characterizations of its ferroelectricity, which is a decisive step for the branch of sliding ferroelectricity.

Finally, although a few molecular ferroelectrics have been developed in recent years, layered molecular ferroelectrics remain rare. Especially, in those few layered molecular ferroelectrics, their spontaneous polarizations are simply induced by the alignment of the discrete organic ammonium cations[21]. And the directions of their polarizations lie in the plane[23]. In this sense, the current work is not a marginal extension of these previous works on molecular ferroelectrics, but host more exotic physical mechanisms of ferroelectricity. The geometric ferroelectricity was once reported only in a few inorganic materials like hexagonal manganites/ferrites[40]. Furthermore, previously studied sliding ferroelectrics, no matter theoretical predictions or experimental reports, are all inorganic materials. Our study extends these two kinds of improper polarity into the field of molecular ferroelectrics, which enriches the scientific connotation of organic-inorganic hybrid systems and broadens the material candidates for future applications.

In summary, we reported systematic experimental characterizations and theoretical scenario of novel ferroelectricity in an amphidynamic crystalline coordination polymer (15-Crown-5)$Cd_3Cl_6$. Our study unambiguously demonstrated its ferroelectricity, which originated from both the geometric and sliding mechanisms. The first contribution is due to the geometric factor of odd number O ions of each crown-ether, and the second one is from the interlayer sliding between layers A and B within one u.c.. These two contributions are coupled and their polarizations are parallel, along the *b*-axis. Our work opens a new era to explore new ferroelectric mechanisms and materials in organic-inorganic hybrid systems.

**Online content**

Any methods, additional references, Nature Research reporting summaries, source data, extended data, supplementary information, acknowledgements, peer review information; details of author contributions and competing interests; and statements of data and code availability are available at https://doi.org/10.1038/xxx.

**METHODS**

**DSC, SHG, and XRD measurements.** Differential scanning calorimetry (DSC) measurements were performed on a NETZSCH DSC (214 Polyma) under nitrogen atmosphere in aluminum crucibles with a heating or cooling rate of 10 K/min. For SHG experiments, samples with particle sizes of 75-150 μm were used to measure the SHG intensity by a pump Nd:YAG laser (1064 nm, 1 Hz repetition rate), and the temperature varies from 220 K to 380 K controlled by a precision temperature controller system (INSTEC Instruments, Model HCS302). Variable-temperature X-ray diffraction analysis was carried out using a Rigaku synergy diffractometer with Mo-K$\alpha$ radiation ($\lambda$ = 0.71073 Å). Data collection, cell refinement, and data reduction were performed using CrysAlisPro (version 1.171.41.112a) XtaLAB Synergy-R online system. The structures were solved by the direct method and refined by the full-matrix method based on $F^2$ using the OLEX2 and SHELXTL (2018) software package. All non-hydrogen atoms were refined anisotropically and the positions of all hydrogen atoms were generated geometrically. The powder XRD pattern (Supplementary Figure 2) was refined using the GSAS Rietveld program[46].

**Dielectric, pyroelectric, and ferroelectric measurements.** For dielectric and ferroelectric measurements, the samples were made with single-crystals cut into the form of thin plates (thickness, ~0.3-0.6 mm) perpendicular to the crystalline *a*-, *b*-, and *c*-axes. The direction of the single crystal was determined by the Rigaku synergy diffractometer [Operating system: CrysAlisPro 1.171.41.112a (Rigaku OD, 2021)]. Silver conduction paste deposited on the plate surfaces was used as the electrodes. Complex dielectric permittivities were measured with a TH2828A impedance analyzer over the frequency range from 500 Hz to 1 MHz with an applied electric field of 0.5 V. Pyroelectric property was measured with an electrometer/high resistance meter (Keithley 6517B) with the heating rate of 10 K/min. $J$-$V$ curves were measured using the double-wave method, which can remove non-hysteresis components in $P$-$E$ loops[47]. Ferroelectric switching measurements were directly carried out on the monocrystal by scanning probe microscopy (SPM) technique through a resonant-enhanced PFM [MFP-3D, Asylum Research, and conductive Pt/Ir-coated silicon probes (EFM-50, Nanoworld)].

**Solid-state NMR experiments.** The $^{13}$C solid-state NMR experiments were performed on a Bruker AVANCE III 400 WB spectrometer operating at 100.06 MHz for $^{13}$C. A 4 mm double resonance MAS probe was used for the $^{13}$C experiments. The temperature-dependent high resolution solid state $^{13}$C NMR spectra was acquired using the single pulse excitation pulse sequence with a recycle delay of 2 s. A 4 mm Magic Angle Spinning (MAS) probe was used in the experiments and the spinning speed was 10 kHz. The $^{13}$C CSA patterns were extracted from the 2D SUPER spectra[48].

**DFT calculation.** The first-principles DFT calculations were performed with the projector augmented wave pseudopotentials as implemented in Vienna *ab initio* Simulation Package (VASP)[49]. The Perdew-Burke-Ernzerhof parameterization of generalized gradient approximation (GGA) was used for the exchange-correlation functional[50]. The plane-wave cutoff energy was 550 eV. The $k$-point grids of 5×3×3 were adopted for both structural relaxation and static computation. To describe the interlayer interaction, the vdW correction of DFT-D3 method is adopted[51], which leads to the lattice constants closer to the experimental value than other corrections. The convergent criterion for the energy was set to $10^{-6}$ eV, and the default criterion of the Hellman-Feynman force during the structural relaxation was <0.01 eV/Å for all atoms. In addition, the polarization was calculated using the standard Berry phase method[52,53]. More benchmarks of DFT calculations can be found in SI (Supplementary Note 3, Fig. 14 and Table 5).

## Data availability

The experimental cif files can be found in CCDC (1875017-1875018 and 2160711-2160716). The experimental and DFT structure files as also uploaded as supplementary files. Other data supporting these findings are available from the corresponding authors upon request.

# Acknowledgments

We thank D.-W. Fu and H.-F. Lu for their suggestions on project conception and structural analysis, and X. Liu, Z. Sheng and M. Liu for their kind help on SHG analysis and Rietveld refinement. Y.Z. acknowledges support from the National Key Research and Development Program of China (grant number 2017YFA0204800) and the Open Project of Shanghai Key Laboratory of Magnetic Resonance (grant number 2018004). S.D. acknowledges support from National Natural Science Foundation of China (grant number 11834002). L.-P.M. acknowledges support from the Jiangxi Provincial Key Laboratory of Functional Molecular Materials Chemistry (grant number 20212BCD42018). Y.-F.Y. acknowledges support from the Xing-Fu-Zhi-Hua Foundation of ECNU. We thank the Big Data Center of Southeast University for providing the facility support on the numerical calculations.

**Author contributions**

Y.Z. and S.D. conceived the project. Y.Z. designed the experiments. S.D. proposed the theoretical mechanisms. L.-P.M. prepared the samples and performed the DSC and SHG measurements. N.W. contributed to PFM measurements. C.S. and H.-Y.Y. contributed to single-crystal measurement and analysis. Y.-F.Y. performed the NMR measurement and analysis. N.D. performed the DFT calculations guided by S.D. L.L. contributed to the analysis of PFM. S.D. and Y.Z. wrote the manuscript, with inputs from all other authors.

**Additional information**

Supplementary Information accompanies this paper at
https://doi.org/10.1038/s41563-022-01322-1

**Competing interests:** The authors declare no competing interests.